\documentclass[a4paper,11pt]{article}
\usepackage{lmodern}

\usepackage[T1]{fontenc}
\usepackage[latin9]{inputenc}
\usepackage{booktabs}
\usepackage{amsmath}
\usepackage{amssymb}

\makeatletter

\providecommand{\tabularnewline}{\\}

\newcommand{\lyxaddress}[1]{
	\par {\raggedright #1
	\vspace{1.4em}
	\noindent\par}
}

\usepackage{mathrsfs}   %\mathscr{L}
\usepackage{slashed}     %\slashed{p}
\usepackage{bbold}  % \mathbb{1} for the identity matrix
\usepackage{url}
\usepackage{graphicx}
\usepackage[colorlinks=true,linkcolor=redLinks,citecolor=greenLinks,urlcolor=redLinks, pdfborder={0 0 1}]{hyperref}
\usepackage{xcolor}
\usepackage{framed}
\usepackage[numbers,sort&compress]{natbib}
\colorlet{shadecolor}{gray!15}

\definecolor{greenLinks}{rgb}{0, 0.6, 0} 
\definecolor{blueLinks}{rgb}{0, 0, 0.6}
\definecolor{redLinks}{rgb}{0.6, 0, 0}
\definecolor{tempText}{rgb}{0.55, 0.10,0.67}
\definecolor{eprintLinks}{rgb}{0.4, 0.4, 0.4}
\definecolor{journalLinks}{rgb}{0.6, 0, 0}

\newcommand{\MYhref}[3][redLinks]{\href{#2}{\color{#1}{#3}}}%

\usepackage{multirow}
\let\orig@Hy@EveryPageAnchor\Hy@EveryPageAnchor
\def\Hy@EveryPageAnchor{%
    \begingroup
    \hypersetup{pdfview=Fit}%
    \orig@Hy@EveryPageAnchor
    \endgroup
}

\let\oldFootnote\footnote
\newcommand\nextToken\relax

\renewcommand\footnote[1]{%
    \oldFootnote{#1}\futurelet\nextToken\isFootnote}

\newcommand\isFootnote{%
    \ifx\footnote\nextToken\textsuperscript{,}\fi}

\definecolor{myPurple}{RGB}{128,0,182}

\usepackage[most]{tcolorbox} 
\definecolor{block-gray}{gray}{0.95}

\newtcolorbox{ToDo}{
    enhanced,
    frame hidden,
    colback=block-gray,
    boxrule=0pt,
    borderline west={0pt}{0pt}{gray!80!black}
}

\newcommand{\WOne}{W}

\makeatother

\begin{document}
\title{A triplet gauge boson with hypercharge one}
\author{Renato M. Fonseca\date{}}
\maketitle

\lyxaddress{\begin{center}
{\Large{}\vspace{-0.5cm}}High Energy Physics Group\\
Departamento de Física Teórica y del Cosmos,\\
Universidad de Granada, E--18071 Granada, Spain\\
~\\
Email: renatofonseca@ugr.es
\par\end{center}}
\begin{abstract}
A vector boson $\WOne_{1}^{\mu}$ with the quantum numbers $\left(\boldsymbol{3},1\right)$
under $SU\left(2\right)_{L}\times U(1)_{Y}$ could in principle couple
with the Higgs field via the renormalizable term $\WOne_{1}^{\mu*}HD_{\mu}H$.
This interaction is known to affect the $T$ parameter and, in so
doing, it could potentially explain the recent CDF measurement of
the W-boson mass.

As it is often the case with vectors, building a viable model with
a $\WOne_{1}$ gauge boson is non-trivial. In this work I will describe
two variations of a minimal setup containing this field; they are
based on an extended $SO(5)\times SU\left(2\right)\times U(1)$ electroweak
group. I will nevertheless show that interactions such as $\WOne_{1}^{\mu*}H\partial_{\mu}H$
are never generated in a Yang-Mills theory. A coupling between $\WOne_{1}$,
$H$ and another Higgs doublet $H^{\prime}$ is possible though.

Finally, I will provide an explicit recipe for the construction of
viable models with gauge bosons in arbitrary representations of the
Standard Model group; depending on the quantum numbers, they may couple
to pairs of Standard Model fermions, or to a Standard Model fermion
and an exotic one.
\end{abstract}

\section{Introduction}

There is abundant speculation on what may lie beyond the Standard
Model (SM). Out of many approaches, there have been attempts to systematically
parameterize the effect of new particles without reference to specific
models. For example, one may look for those fields which can interact
with Standard Model particles --- such as a pair of fermions ---
in a renormalizable way \cite{delAguila:2010mx,deBlas:2017xtg}. At
low energies, they give rise to four-fermion interactions much like
those in the Fermi theory, which led to the discovery of weak interactions.

A field which has received fairly little attention is $\WOne_{1}$,
with the $SU\left(2\right)_{L}\times U(1)_{Y}$ quantum numbers $\left(\boldsymbol{3},1\right)$
and transforming as a 4-vector under the Lorentz group. It can couple
only to the Higgs doublet, doing so via the interaction \cite{delAguila:2010mx,deBlas:2017xtg,DiLuzio:2022xns}
\begin{equation}
\frac{\kappa}{2}\WOne_{1}^{\mu,a*}H^{T}\left(i\sigma_{2}\sigma_{a}\right)D_{\mu}H\,.\label{eq:W1-interaction}
\end{equation}
Once integrated out, the fermionphobic $\WOne_{1}^{\mu}$ gives rise
to the effective dimension six interaction
\begin{gather}
-\frac{\left|\kappa\right|^{2}}{4m_{\WOne_{1}}^{2}}\left(\mathcal{O}_{H}^{1}+\mathcal{O}_{H}^{3}\right)\label{eq:dim-6-ops}
\end{gather}
with $\mathcal{O}_{H}^{1}\equiv\left[\left(D^{\mu}H\right)^{\dagger}\left(D_{\mu}H\right)\right]\left(H^{\dagger}H\right)$
and $\mathcal{O}_{H}^{3}\equiv\left|H^{\dagger}\left(D^{\mu}H\right)\right|^{2}$.
In turn, this last operator contributes to the $\widehat{T}=\alpha T$
parameter \cite{Peskin:1990zt,Peskin:1991sw,Barbieri:2004qk} as follows:
\begin{equation}
\widehat{T}\equiv\frac{\Pi_{W^{3}W^{3}}\left(0\right)-\Pi_{W^{+}W^{-}}\left(0\right)}{m_{W}^{2}}=\frac{\left|\kappa\right|^{2}}{4}\frac{v^{2}}{m_{\WOne_{1}}^{2}}\,;\;v^{2}\equiv\left\langle H^{\dagger}H\right\rangle \approx\left(174\textrm{ GeV}\right)^{2}\,.
\end{equation}

Recently, the CDF collaboration published a surprising large value
of the $W$ mass \cite{CDF:2022hxs}, 
\begin{equation}
m_{W}=80433.5\pm9.4\textrm{ MeV},
\end{equation}
which seems to be well fitted by $\widehat{T}\approx\left(8.8\pm1.4\right)\times10^{-4}$
\cite{Strumia:2022qkt} (see also \cite{Asadi:2022xiy}). This would
correspond to a value of $m_{\WOne_{1}}/\left|\kappa\right|\approx2.9\textrm{ TeV}$
\cite{DiLuzio:2022xns}, which is not ruled out by LHC searches.
The authors of \cite{Bagnaschi:2022whn} find that $\WOne_{1}$ is
one of the single-field extensions of the Standard Model which best
fits the CDF data, adding nonetheless that this field is not commonly
found in unified gauge theories. Notice that unlike a $Z^{\prime}$
or a $W^{\prime}$, a single $\WOne_{1}$ cannot be produced in a
proton-proton collision, neither through vector-boson fusion nor through
the Drell-Yan process. For this reason, searches for new vector fields
by ATLAS and CMS \cite{ATLAS:2018sbw,ATLAS:2020qiz,CMS:2021fyk,CMS:2021xor}
--- which in some cases exclude masses up to 5 TeV --- do not apply
to $\WOne_{1}$. On the other hand, mass limits from pair production
searches reach at most 1 to 2 TeV (see for instance \cite{ATLAS:2021oiz,CMS:2020wzx})
for colored fields. For uncolored ones, the LHC reach is even lower,
as can be seen from the limits on the scalar analog of $\WOne_{1}$,
which is present in seesaw type-II mechanism for neutrino masses \cite{ATLAS:2018ceg,CMS:2012dun}.

It is far from clear that this new determination of the $W$ mass
will resist the test of time, and in fact the CDF measurement is in
tension with other direct and indirect determinations of $m_{W}$
\cite{deBlas:2022hdk,Bagnaschi:2022whn,Lu:2022bgw}. Nonetheless ---
independently of the validity of this result --- it is worth considering
in detail the phenomenology associated to a $\WOne_{1}$ vector, which
will inevitably depend on the ultraviolet origin of the field. To
my knowledge, there have been no previous attempts of incorporating
this vector in a complete model. With that in mind, in this work I
will present a minimal setup where $\WOne_{1}$ is a gauge boson associated
to an extended electroweak group, which is spontaneously broken at
the TeV scale or above. Another possibility, not considered here,
is that $\WOne_{1}$ is a composite field rather than a fundamental
one.\\
~

It turns out that the trilinear interaction between a gauge field
 and two scalars, $\phi$ and $\phi^{\prime}$, must be anti-symmetric
under the exchange of the scalars. This means that for $\phi=\phi^{\prime}=H$,
which is an $SU(2)$ doublet, there can be no coupling to the triplet
$\WOne_{1}$. I will discuss this point in section \ref{sec2:no-coupling},
after which I will detail a realistic model for $\WOne_{1}$ based
on the gauge group $SO(5)\times SU\left(2\right)\times U(1)$ (section
\ref{sec3:A-minimal-model}). In it the new vector does not couple
to matter; however, one can modify the setup so that it does (section
\ref{sec4:An-alternative:-charging}).

Looking beyond $\WOne_{1}$, it is not straightforward to build Yang-Mills
theories with gauge bosons transforming according to random representations
of the Standard Model group. In many cases, there might be no model
in the literature containing such vector fields. To improve on this
situation, in section \ref{sec5:Arbitrary-gauge-bosons} I describe
an explicit recipe for building models with gauge bosons assigned
to arbitrary representations of the Standard Model group. Section
\ref{sec:Final-comments} summarizes the main conclusions in this
work.

\section{\label{sec2:no-coupling}$\WOne_{1}^{\mu*}HD_{\mu}H$ and similar
couplings in a Yang-Mills theory}

In order for a gauge boson to potentially couple to the combination
$HD_{\mu}H$, rather than to $H^{*}D_{\mu}H$, it must be that the
Higgs is a linear combination of at least two fields --- let us call
them $H_{u}=\left(\boldsymbol{2},1/2\right)$ and $H_{d}=\left(\boldsymbol{2},-1/2\right)$
--- which are part of the same irreducible representation $\Omega$
of the gauge group. Then, from the kinetic term of this last field,
one might hope to get the sought after interaction:
\begin{equation}
\left(D^{\mu}\Omega\right)^{\dagger}\left(D_{\mu}\Omega\right)\propto\cdots+\left(\WOne_{1}^{\mu*}H_{d}^{*}D_{\mu}H_{u}+\textrm{h.c.}\right)\propto\cdots+\left(\WOne_{1}^{\mu*}HD_{\mu}H+\textrm{h.c.}\right)\,.\label{eq:kinetic-term}
\end{equation}
I have used here the proportionality sign to avoid distractions with
the prefactors of each expression; likewise I also did not track carefully
how the $SU(2)$ indices are contracted. However, by tracking attentively
relative signs, the reader will see that while one can indeed achieve
a $\WOne_{1}^{\mu*}H_{d}^{*}D_{\mu}H_{u}$ coupling which goes on
to contribute to the interaction $\WOne_{1}^{\mu*}HD_{\mu}H$, we
must also take into account a term $\WOne_{1}^{\mu*}H_{u}D_{\mu}H_{d}^{*}$
with the opposite effect. All in all, this means that the prefactor
of the Higgs-Higgs-$\WOne_{1}$ coupling is null.

Perhaps --- one might think --- this cancellation is specific to
the minimalist scalar setup described above. That is not true: I will
argue in the following that in a Yang-Mills theory an interaction
$\mathcal{A}^{\mu}\phi\phi^{\prime}$ between a gauge boson $\mathcal{A}^{\mu}$
and two scalars $\phi$ and $\phi^{\prime}$, through a derivative,
must be anti-symmetric under an exchange of these two scalars. Since
in our particular example the $SU(2)$ quantum numbers force the two
$H$'s to be contracted symmetrically with the triplet $\WOne_{1}$,
it must be that the coefficient of the interaction is zero, regardless
of the details of the model.

To see this, we may start by decomposing all scalars in a model in
real components, and collecting them in a column vector $\Phi$ ($=\Phi^{*}$).
Any gauge transformation can be represented through a matrix $\exp\left(i\varepsilon_{a}T_{a}\right)$
which must be both real and unitary, hence
\begin{equation}
T_{a}=T_{a}^{\dagger}=-T_{a}^{*}\,.
\end{equation}
These anti-symmetric $T_{a}$ generators regulate the interaction\footnote{There is more than one coupling constant $g$ if the gauge group is
semi-simple. Nonetheless, such complication is of no consequence to
the present discussion.}
\begin{equation}
\Phi^{T}\left(igT_{a}\mathcal{A}_{a}^{\mu}\right)\left(D_{\mu}\Phi\right)
\end{equation}
obtained from the kinetic term $\left(D^{\mu}\Phi\right)^{T}\left(D_{\mu}\Phi\right)/2$,
so in any other basis (such as the electroweak one, or perhaps the
mass basis), with $\Phi=U\Phi^{\prime}$, the all-important matrices
$igU^{T}T_{a}U$ remains anti-symmetric. Therefore, for any pair of
irreducible representations $\phi$ and $\phi^{\prime}$ of the gauge
group, we extract from the off-diagonal part of $igU^{T}T_{a}U$ a
term
\begin{equation}
\mathcal{A}_{a}^{\mu}\left[\phi^{T}X_{a}\left(D_{\mu}\phi^{\prime}\right)-\phi^{\prime T}X_{a}\left(D_{\mu}\phi\right)\right]
\end{equation}
for some real matrices $X_{a}$. In the particular case when $\phi=\phi^{\prime}$
we may write the interaction as
\begin{equation}
\mathcal{A}_{a}^{\mu}\left[\phi^{T}X_{a}\left(D_{\mu}\phi\right)\right]\,\textrm{ with }X_{a}=-X_{a}^{T}\,;
\end{equation}
the $X_{a}$ are nothing but diagonal blocks of the bigger $igU^{T}T_{a}U$
matrices mentioned above. The last expression shows explicitly that
the gauge indices of the two $\phi$'s must contract anti-symmetrically:
a coupling $B_{1}^{\mu*}HD_{\mu}H$ with the field $B_{1}=\left(\boldsymbol{1},1\right)$
is fine, given that two doublets contract anti-symmetrically to form
a singlet; by an analogous argument, a coupling $\WOne_{1}^{\mu*}HD_{\mu}H$
is not.

The anti-symmetry of an interaction between $\mathcal{A}^{\mu}$ and
two scalars was just derived in the context of a Yang-Mills theory.
But for the sake of argument, let us consider adding a symmetric part
to the $X_{a}$ matrices (keeping $\mathcal{A}^{\mu}$ massless).
The Feynman rule for the vertex $\phi_{i}-\phi_{j}-\mathcal{A}_{a}^{\mu}$
is given by the expression
\begin{equation}
\left(X_{a}\right)_{ij}p_{\phi_{j}}^{\mu}+\left(X_{a}\right)_{ji}p_{\phi_{i}}^{\mu}=\left[\left(X_{a}\right)_{ij}-\left(X_{a}\right)_{ji}\right]p_{\phi_{j}}^{\mu}-\left(X_{a}\right)_{ji}p_{\mathcal{A}}^{\mu}
\end{equation}
where all momenta are assumed to be incoming, so that $p_{\phi_{i}}^{\mu}+p_{\phi_{j}}^{\mu}+p_{\mathcal{A}}^{\mu}=0$.
The $p_{\mathcal{A}}^{\mu}$ term is irrelevant when this vertex is
contracted with a polarization vector $\epsilon_{\mu}$ (since $\epsilon_{\mu}p_{\mathcal{A}}^{\mu}=0$).
On the other hand, contracting this same term with a $\mathcal{A}_{a}^{\mu}$
propagator in the $R_{\xi}$ gauge yields an unphysical contribution
proportional to the $\xi$ parameter. So the $p_{\mathcal{A}}^{\mu}$
term can be dropped altogether, and with it the symmetric part of
$X_{a}$ disappears from the vertex expression. At the Lagrangian
level, this can be traced back to the fact that the scalar interaction
with the symmetric part of $X_{a}$ can be swapped by a total derivative
and a term proportional to $\partial_{\mu}\mathcal{A}_{a}^{\mu}$,
which does not affect physically relevant calculations:\footnote{For a broad class of functions $f_{a}$, meaningful predictions remain
unchanged if we add to the Lagrangian a term $f_{a}\left(A,\textrm{other fields}\right)^{2}/2\xi$.
This is the well known gauge-fixing term of Yang-Mills theories. For
real operators $\mathcal{O}_{a}$ (such as $\Phi^{T}X_{a}\Phi$) and
some number $\alpha$ we may pick $f_{a}=\partial_{\mu}\mathcal{A}_{a}^{\mu}+\alpha\xi\mathcal{O}_{a}$,
in which case we conclude that
\[
\frac{1}{2\xi}\left(\partial_{\mu}\mathcal{A}_{a}^{\mu}\right)^{2}+\alpha\left(\partial_{\mu}\mathcal{A}_{a}^{\mu}\right)\mathcal{O}_{a}+\frac{\alpha\xi^{2}}{2}\mathcal{O}_{a}^{2}
\]
is irrelevant for any value of $\xi$ and $\alpha$. As a consequence,
in the limit $\xi\rightarrow0$ (associated to the Lorenz gauge) the
operators proportional to $\partial_{\mu}\mathcal{A}_{a}^{\mu}$ can
be dropped from the Lagrangian.}
\begin{equation}
\partial_{\mu}\left(\mathcal{A}_{a}^{\mu}\phi^{T}X_{a}\phi\right)=\left(\partial_{\mu}\mathcal{A}_{a}^{\mu}\right)\phi^{T}X_{a}\phi+\mathcal{A}_{a}^{\mu}\phi^{T}\left(X_{a}+X_{a}^{T}\right)\left(\partial_{\mu}\phi\right)\,.
\end{equation}

Finally, let us consider the spinor-helicity formalism, where it is
straightforward to compute the scattering amplitude for a spin 1 field
(particle \#1) and two scalars (particles \#2 and \#3) when all three
are massless. Poincaré invariance and unitarity are the only extra
assumptions. In the widely used bracket notation, depending on the
helicity of the spin 1 particle, the amplitude is proportional to
either
\begin{equation}
\frac{\left\langle 12\right\rangle \left\langle 31\right\rangle }{\left\langle 23\right\rangle }\textrm{ or }\frac{\left[12\right]\left[31\right]}{\left[23\right]}\,.
\end{equation}
In both cases, permuting the two scalars ($2\leftrightarrow3$) yields
a minus sign so --- at least for massless scalars --- the spinor-helicity
formalism corroborates the anti-symmetry of a $\mathcal{A}^{\mu}\phi\phi^{\prime}$
interaction.

~

In conclusion, a fundamental field $\WOne_{1}$ which gets a mass
through the Higgs mechanism cannot have a $\WOne_{1}^{\mu*}HD_{\mu}H$
coupling; such an interaction is absent if $\WOne_{1}$ is a gauge
boson.

With this advance warning, I will proceed to describe the basic features
of a minimal extension of the Standard Model where this field appears.

\section{\label{sec3:A-minimal-model}A model for $\WOne_{1}^{\mu}$}

Since $\WOne_{1}$ is charged under both $SU\left(2\right)_{L}$ and
$U(1)_{Y}$, this field as well as the Standard Model $W$ and $B$
must be gauge bosons associated to some group which includes the
electroweak one, $SU\left(2\right)_{L}\times U(1)_{Y}$. The adjoint
representation of such a group must be of size at least $6+3+1=10$,
considering that this is the total number of real field components
in $\WOne_{1}$, $W$ and $B$. It turns out that the adjoint representation
of the group $SO(5)$ (whose algebra is isomorphic to $Sp(4)$) is
precisely 10 dimensional. Furthermore, $SU(2)\times U(1)$ is a subgroup
of $SO(5)$, and under it the spinor representation branches as follows:\footnote{In fact, $SO(5)$ has two inequivalent $SU(2)\times U(1)$ subgroups.
The other embedding is associated to the branching rule $\boldsymbol{4}\rightarrow\left(\boldsymbol{1},1/2\right)+\left(\boldsymbol{1},-1/2\right)+\left(\boldsymbol{2},0\right)$,
which is not relevant for the present work.}
\begin{equation}
\boldsymbol{4}\rightarrow\left(\boldsymbol{2},1/2\right)+\left(\boldsymbol{2},-1/2\right)\,.
\end{equation}
The adjoint decomposes in the manner alluded above, namely
\begin{equation}
\boldsymbol{10}\rightarrow\left(\boldsymbol{1},0\right)+\left(\boldsymbol{3},0\right)+\underbrace{\left(\boldsymbol{3},1\right)+\left(\boldsymbol{3},-1\right)}_{\WOne_{1}}\,.\label{eq:10}
\end{equation}
Note that the decomposition of the spinor representation implies that
a scalar field transforming as a $\boldsymbol{4}$ contains both an
$H_{u}$- and an $H_{d}$-like field, which is precisely what one
needs to generate a $H-H^{\prime}-\WOne_{1}^{\mu}$ coupling, as discussed
in the previous section.

So far everything looks promising. Nevertheless, when it comes to
fermions, it is challenging to charge them non-trivially under $SO(5)$.
For example, the left-handed leptons $L=\left(\boldsymbol{2},-1/2\right)$
necessarily interact via the $\WOne_{1}$ gauge bosons with fermions
whose charges are $\left(\boldsymbol{2},-3/2\right)$, $\left(\boldsymbol{2},1/2\right)$,
$\left(\boldsymbol{4},-3/2\right)$, or $\left(\boldsymbol{4},1/2\right)$,
none of which are part of the Standard Model. The same thing happens
with quarks, and therefore one would need to find vector-like partners
for these new fields in order to give them masses above those of $\left\langle H\right\rangle \approx174$
GeV. With the help of an extra $U(1)$ which would provide more flexibility
in forming the SM hypercharge group, one can certainly find $SO(5)\times U(1)$
representations with the sought-after fermions, however these tend
to propagate the problem by introducing further chiral states.

We are therefore guided to the possibility that no chiral fermion
is charged under $SO(5)$, and instead the full electroweak group
is $SO(5)\times SU(2)\times U(1)$, which contains $SU(2)^{\prime}\times U(1)^{\prime}\times SU(2)\times U(1)$;
in turn, its diagonal subgroup is $SU\left(2\right)_{L}\times U(1)_{Y}$:

\begin{equation}
SO(5)\times SU(2)\times U(1)\rightarrow\underbrace{SU(2)^{\prime}\times SU(2)}_{\supset SU\left(2\right)_{L}}\times\underbrace{U(1)^{\prime}\times U(1)}_{\supset U(1)_{Y}}\,.\label{eq:group-breaking}
\end{equation}
The adjoint representation of the extended electroweak group includes
the $SU\left(2\right)_{L}\times U(1)_{Y}$ representations $B,Z^{\prime}=\left(\boldsymbol{1},0\right)$,
$W,W^{\prime}=\left(\boldsymbol{3},0\right)$ and $\WOne_{1}=\left(\boldsymbol{3},1\right)$.
These gauge bosons acquire a mass proportional to the $SO(5)$-symmetry
breaking scale, except for the Standard Model $B$ and $W$, which
are less massive. The latter fields are a mixture of the $SO(5)$
gauge bosons with those of $SU(2)\times U(1)$.

Note also that a scalar $\Omega$ transforming as a spinor under $SO(5)$
would couple to $\WOne_{1}$, but not to fermions since they are uncharged
under this group. Consequently, in order to have Yukawa interactions
one also needs an $SO(5)$-singlet scalar $\widehat{H}$. Finally,
the extended electroweak group can be broken to $SU\left(2\right)_{L}\times U(1)_{Y}$
with a non-zero vacuum expectation value (VEV) of some field $\chi$.

~

With the above general considerations, we are in a position to flesh
out a model. The requirements discussed earlier on the scalar sector
are fully met by the fields in table \ref{tab:scalars}. In terms
of Standard Model $SU\left(2\right)_{L}\times U(1)_{Y}$ representations,
$\chi$ contains a component with quantum numbers $\left(\boldsymbol{1},0\right)$;
as we shall see, its VEV preserves only the subgroup $SU\left(2\right)_{L}\times U(1)_{Y}$
of $SO(5)\times SU(2)\times U(1)$. Note also that there is a total
of 3 Higgs doublets in $\Omega$ and $\widehat{H}$, which can mix
to produce the Standard Model field $H$.

\begin{table}
\begin{centering}
\begin{tabular}{ccc}
\toprule 
Scalar & $SO(5)\times SU(2)\times U(1)$ & $SU(2)_{L}\times U(1)_{Y}$ decomposition\tabularnewline
\midrule
$\Omega$ & $\left(\boldsymbol{4},\boldsymbol{1},0\right)$ & $\left(\boldsymbol{2},-\frac{1}{2}\right)+\left(\boldsymbol{2},\frac{1}{2}\right)$\tabularnewline
$\widehat{H}$ & $\left(\boldsymbol{1},\boldsymbol{2},\frac{1}{2}\right)$ & $\left(\boldsymbol{2},\frac{1}{2}\right)$\tabularnewline
$\chi$ & $\left(\boldsymbol{4},\boldsymbol{2},\frac{1}{2}\right)$ & $\left(\boldsymbol{1},0\right)+\left(\boldsymbol{1},1\right)+\left(\boldsymbol{3},0\right)+\left(\boldsymbol{3},1\right)$\tabularnewline
\bottomrule
\end{tabular}
\par\end{centering}
\caption{\label{tab:scalars}The quantum numbers of the three scalars in the
model, under the extended electroweak group $SO(5)\times SU(2)\times U(1)$.
A non-zero vaccum expectation value of $\chi$ can break this symmetry
down to $SU(2)_{L}\times U(1)_{Y}$. The transformation properties
of the scalars under this latter group are shown in the last column.
All fermions transform trivially under $SO(5)$.}
\end{table}

Let us now establish a set of generators of the 4-dimensional representation
of $SO(5)$. Notice again that this group is isomorphic to $Sp(4)$,
which can be defined via the set of 4-dimensional matrices $G$ satisfying
the relation $G^{T}JG=J$; $J$ is a non-singular anti-symmetric matrix
which is often taken to have the block form
\begin{equation}
J=\left(\begin{array}{cc}
\boldsymbol{0} & \boldsymbol{1}_{2}\\
-\boldsymbol{1}_{2} & \boldsymbol{0}
\end{array}\right)\,.
\end{equation}
Rewriting $G$ as $\exp\left(i\varepsilon^{a}T^{a}\right)$ with real
$\varepsilon^{a}$ parameters --- and requiring also that $G^{\dagger}G=\boldsymbol{1}_{4}$
--- leads to infinitesimal generators of the form 
\begin{equation}
\varepsilon_{a}T^{a}=\left(\begin{array}{cc}
\boldsymbol{B} & \boldsymbol{C}\\
\boldsymbol{C^{*}} & \boldsymbol{-B}^{*}
\end{array}\right)
\end{equation}
where $\boldsymbol{B}$ and $\boldsymbol{C}$ are arbitrary 2-dimensional
hermitian and symmetric matrices, respectively. However, it is more
convenient to work on a basis where the last two entries of the 4-dimensional
space are rotated with the $\epsilon=i\sigma_{2}$ matrix, such that
\begin{equation}
J=\left(\begin{array}{cc}
\boldsymbol{0} & \epsilon\\
\epsilon & \boldsymbol{0}
\end{array}\right)\,.
\end{equation}
The 10 generators can then be picked to have the following form:\footnote{This is $\sqrt{2}$ times the matrices given by \texttt{RepMatrices{[}SO5,4{]}}
in \texttt{GroupMath} \cite{Fonseca:2020vke}, after a reordering.}
\begin{equation}
\varepsilon^{a}T^{a}\equiv\frac{1}{2}\left(\begin{array}{cccc}
\varepsilon_{3}+\varepsilon_{10} & \varepsilon_{1}-i\varepsilon_{2} & \varepsilon_{8}-i\varepsilon_{5} & \sqrt{2}(\varepsilon_{6}-i\varepsilon_{9})\\
\varepsilon_{1}+i\varepsilon_{2} & -\varepsilon_{3}+\varepsilon_{10} & \sqrt{2}(\varepsilon_{4}-i\varepsilon_{7}) & -\varepsilon_{8}+i\varepsilon_{5}\\
\varepsilon_{8}+i\varepsilon_{5} & \sqrt{2}(\varepsilon_{4}+i\varepsilon_{7}) & \varepsilon_{3}-\varepsilon_{10} & \varepsilon_{1}-i\varepsilon_{2}\\
\sqrt{2}(\varepsilon_{6}+i\varepsilon_{9}) & -\varepsilon_{8}-i\varepsilon_{5} & \varepsilon_{1}+i\varepsilon_{2} & -\varepsilon_{3}-\varepsilon_{10}
\end{array}\right)\,.
\end{equation}
If we throw away $\varepsilon_{4,\cdots,9}$, this becomes a block-diagonal
matrix and in fact $T^{1,2,3}$ are generators of an important $SU(2)^{\prime}$
subgroup (see expression (\ref{eq:group-breaking})), with Pauli matrices
on the diagonal blocks; $T^{10}$ generates a $U(1)^{\prime}$ which
commutes with this $SU(2)^{\prime}$.

The scalar $\chi=\left(\boldsymbol{4},\boldsymbol{2},\frac{1}{2}\right)$
can be seen as a two index field $\chi_{jk}$ with $SO(5)$ acting
on $j$ and $SU(2)$ on $k$. With this understanding,
\begin{align}
D_{\mu}\chi_{jk} & =\partial_{\mu}\chi_{jk}+ig_{A}A_{\mu}^{a,SO(5)}T_{jj^{\prime}}^{a}\chi_{j^{\prime}k}+\frac{i}{2}g_{B}A_{\mu}^{b,SU(2)}\sigma_{kk^{\prime}}^{b}\chi_{jk^{\prime}}+\frac{1}{2}g_{C}A_{\mu}^{U(1)}\chi_{jk}\,.
\end{align}
The VEV
\begin{equation}
\left\langle \chi\right\rangle \propto\left(\begin{array}{cc}
0 & 0\\
0 & 0\\
0 & -1\\
1 & 0
\end{array}\right)
\end{equation}
breaks all but 4 linear combinations of the original 14 generators
of the group $SO(5)\times SU(2)\times U(1)$, namely
\begin{equation}
\left(T_{jj^{\prime}}^{1,2,3}\delta_{kk^{\prime}}+\frac{1}{2}\delta_{jj^{\prime}}\sigma_{kk^{\prime}}^{1,2,3}\right)\left\langle \chi\right\rangle _{j^{\prime}k^{\prime}}=\left(T_{jj^{\prime}}^{10}\delta_{kk^{\prime}}+\frac{1}{2}\delta_{jj^{\prime}}\delta_{kk^{\prime}}\right)\left\langle \chi\right\rangle _{j^{\prime}k^{\prime}}=0\,.
\end{equation}
They generate the $SU(2)_{L}$ diagonal subgroup of $SU(2)\times SU(2)^{\prime}$,
together with $U(1)_{Y}$ which is the diagonal subgroup of $U(1)\times U(1)^{\prime}$.
It follows directly from these relations that the Standard Model gauge
couplings $g$ and $g^{\prime}$ are given by the expressions
\begin{align}
g^{-2} & =g_{A}^{-2}+g_{B}^{-2}\,,\label{eq:g2}\\
\left(g^{\prime}\right)^{-2} & =g_{A}^{-2}+g_{C}^{-2}\,.\label{eq:gp2}
\end{align}

We can also extract from $\left(D_{\mu}\left\langle \chi\right\rangle _{jk}\right)^{*}D_{\mu}\left\langle \chi\right\rangle _{jk}$
the leading contribution to the various gauge boson masses (the Higgs
doublet VEVs produce corrections):
\begin{align}
m_{\WOne_{1}}^{2} & =g_{A}^{2}\left\langle \chi\right\rangle ^{2}\,,\\
m_{W^{\prime}}^{2} & =\left(g_{A}^{2}+g_{B}^{2}\right)\left\langle \chi\right\rangle ^{2}\,,\\
m_{Z^{\prime}}^{2} & =\left(g_{A}^{2}+g_{C}^{2}\right)\left\langle \chi\right\rangle ^{2}\,,
\end{align}
where $\left\langle \chi\right\rangle ^{2}\equiv\left\langle \chi\right\rangle _{jk}^{*}\left\langle \chi\right\rangle _{jk}$.
From equations (\ref{eq:g2}) and (\ref{eq:gp2}) plus the known values
of couplings at the electroweak scale ($g\approx0.65$ and $g^{\prime}\approx0.36$)
it must be that $g_{C}$ is smaller than $g_{B}$, so
\begin{equation}
m_{\WOne_{1}}<m_{Z^{\prime}}<m_{W^{\prime}}\,.
\end{equation}
Furthermore, the two independent ratios which can be computed out
of the three masses are related by the expression
\begin{equation}
\frac{m_{\WOne_{1}}^{2}}{m_{Z^{\prime}}^{2}}\left(\frac{m_{W^{\prime}}^{2}}{m_{Z^{\prime}}^{2}}-\tan^{2}\theta_{w}\right)=\left(1-\tan^{2}\theta_{w}\right)\frac{m_{W^{\prime}}^{2}}{m_{Z^{\prime}}^{2}}
\end{equation}
where $\tan^{2}\theta_{w}\equiv g^{\prime2}/g^{2}\approx0.30$. Notice
that while $m_{W^{\prime}}/m_{\WOne_{1}}$ can be arbitrarily large,
$m_{Z^{\prime}}/m_{\WOne_{1}}$ is bounded between 1 and $\sqrt{1/\left(1-\tan^{2}\theta_{w}\right)}\approx1.19$.

~

Besides interacting with other gauge bosons, $\WOne_{1}$ couples
to scalars. In particular $\Omega$ contains an up- plus a down-type
Higgs doublet, $H_{u}=\left(\boldsymbol{2},1/2\right)$ and $H_{d}=\left(\boldsymbol{2},-1/2\right)$,
so --- as foreseen in section \ref{sec2:no-coupling} --- from the
covariant derivative of $\Omega$ we get the interactions
\begin{equation}
\frac{g_{A}}{\sqrt{2}}\WOne_{1}^{\mu,a*}\left[H_{d}^{\dagger}\sigma_{a}\left(D_{\mu}H_{u}\right)-\left(D_{\mu}H_{d}\right)^{\dagger}\sigma_{a}H_{u}\right]+\textrm{h.c..}
\end{equation}
The fields $H_{u}$, $H_{d}^{*}$ and $\widehat{H}$ mix, generating
the 125 GeV scalar $H$ of the Standard Model, as well as two heavier
doublets: $H^{\prime}$ and $H^{\prime\prime}$. No matter what is
the form of this mixing, there will be no $HH$, $H^{\prime}H^{\prime}$
or $H^{\prime\prime}H^{\prime\prime}$ interaction with $\WOne_{1}$.
However, we do get the term
\begin{equation}
\kappa_{HH^{\prime}}\WOne_{1}^{\mu,a*}\left[H^{T}\left(i\sigma_{2}\sigma_{a}\right)D_{\mu}H^{\prime}-H^{\prime T}\left(i\sigma_{2}\sigma_{a}\right)D_{\mu}H\right]
\end{equation}
plus similar ones for other combinations of the three scalar doublets.

In this model, fermions do not couple to $\WOne_{1}$. They also do
not not couple to the $\chi$ scalar, which is significant because
this fields contains an $\left(\boldsymbol{3},1\right)$ representation.
These are the quantum numbers of the mediator in the type-II seesaw
mechanism, capable of generating neutrino masses. But for that to
happen, $\chi$ would need to interact with leptons, which is not
the case. Note also that the model conserves lepton number, hence
neutrinos are massless. This changes, for example, with the introduction
of an extra scalar with the quantum numbers $\left(\boldsymbol{1},\boldsymbol{3},1\right)$
under the extended electroweak group.

\section{\label{sec4:An-alternative:-charging}An alternative: charging fermions
under $SO(5)$}

No fermion is charged under $SO(5)$ in the model described above.
However, on top of the chiral ones, we may introduce vector-like fermions
which do couple to the gauge bosons of this group, without producing
dangerous light fields. One possibility is this: for every chiral
fermion $F=Q,u^{c},d^{c},L,e^{c}$ with the quantum numbers $\left(\boldsymbol{C},\boldsymbol{1},\boldsymbol{L},y\right)$
under $SU(3)_{C}\times SO(5)\times SU(2)\times U(1)$, we introduce
the vector-like pair of left-handed Weyl spinors $\left(\mathbb{F},\overline{\mathbb{F}}\right)$
with $\mathbb{F}=\left(\boldsymbol{C},\boldsymbol{4},\boldsymbol{L},y\right)$
and its conjugate representation $\overline{\mathbb{F}}=\left(\overline{\boldsymbol{C}},\boldsymbol{4},\boldsymbol{L},-y\right)$.\footnote{The spinor representation of $SO(5)$ is pseudo-real, therefore it
is isomorphic to its conjugate. The same is true for any representations
of $SU(2)$.} In other words, apart from transforming as a spinor of $SO(5)$,
all other quantum numbers of $\mathbb{F}$ are the same as those of
$F$. 
\begin{table}
\begin{centering}
\begin{tabular}{ccc}
\toprule 
Field & Spin & $SU(3)_{C}\times SO(5)\times SU(2)\times U(1)$\tabularnewline
\midrule
$F=Q,u^{c},d^{c},L,e^{c}$ & 1/2 & As in the SM; $\boldsymbol{1}$ under $SO(5)$\tabularnewline
$\mathbb{F}=\mathbb{Q},\mathbb{u}^{c},\mathbb{d}^{c},\mathbb{L},\mathbb{e}^{c}$ & 1/2 & As in the SM; $\boldsymbol{4}$ under $SO(5)$\tabularnewline
$\overline{\mathbb{F}}=\overline{\mathbb{Q}},\overline{\mathbb{u}^{c}},\overline{\mathbb{d}^{c}},\overline{\mathbb{L}},\overline{\mathbb{e}^{c}}$ & 1/2 & Complex conjugate of $\mathbb{F}$\tabularnewline
$\Omega$ & 0 & $\left(\boldsymbol{4},\boldsymbol{1},0\right)$\tabularnewline
$\chi$ & 0 & $\left(\boldsymbol{4},\boldsymbol{2},\frac{1}{2}\right)$\tabularnewline
\bottomrule
\end{tabular}
\par\end{centering}
\caption{\label{tab:model-2}Field content of the second model, containing
vector-like fermions $\left(\mathbb{F},\overline{\mathbb{F}}\right)$
unlike the first setup. Furthermore, adequate Yukawa interactions
can be achieved without the scalar $\widehat{H}$ (compare with table
\ref{tab:scalars}).}
\end{table}
The scalar $\chi$, as before, is needed for symmetry breaking; it
also participates in Yukawa interactions with the new fermions. Between
the two remaining scalars ($\widehat{H}$ and $\Omega$) there is
need for just one: I'll keep $\Omega$ (see table \ref{tab:model-2}).

With these charge assignments, we may have the following masses and
interactions:
\begin{equation}
\sum_{F}\left(y_{F}^{\Omega}F\overline{\mathbb{F}}\Omega+m_{F}\mathbb{F}\overline{\mathbb{F}}\right)+y_{QU}^{\chi}Q\mathbb{u}^{c}\chi+y_{QU}^{\chi\prime}\mathbb{Q}u^{c}\chi+y_{QD}^{\chi}Q\mathbb{d}^{c}\chi^{*}+y_{QD}^{\chi\prime}\mathbb{Q}d^{c}\chi^{*}+y_{LE}^{\chi}L\mathbb{e}^{c}\chi^{*}+y_{LE}^{\chi\prime}\mathbb{L}e^{c}\chi^{*}+\textrm{h.c.}\,.\label{eq:yukawa-interactions-extra}
\end{equation}
Note that $\Omega$ needs to be a complex field, so terms of the form
$F\overline{\mathbb{F}}\Omega^{*}$ are also allowed. However, for
simplicity, I will consider that they has been removed (with a $Z_{4}$
symmetry, for example).\footnote{The charges $i$, 1, $-i$, 1, 1, 1, $i$, $-i$, 1, $i$, $-i$,
1, $i$, $-i$, $-i$, 1, 1 for the fields $\Omega$, $\chi$, $Q$,
$\mathbb{Q}$, $\overline{\mathbb{Q}}$, $u^{c}$, $\mathbb{u}^{c}$,
$\overline{\mathbb{u}^{c}}$, $d^{c}$, $\mathbb{d}^{c}$, $\overline{\mathbb{d}^{c}}$,
$L$, $\mathbb{L}$, $\overline{\mathbb{L}}$, $e^{c}$, $\mathbb{e}^{c}$,
$\overline{\mathbb{e}^{c}}$ successfully achieve the goal.}

Let us now consider what happens under the Standard Model subgroup.
The $F$'s don't transform under $SO(5)$, so for convenience one
may use the same name, $F$, to designate their quantum numbers under
the reduced $SU(2)\times U(1)$ group. With this understanding, a
quick way of grasping all the fermion sub-representation in $\mathbb{F}/\overline{\mathbb{F}}$
is to note that $\Omega=H_{u}+H_{d}$, so $\overline{\mathbb{F}}$
contains all the states which couple to the product $FH_{u}$ as well
as all those which couple to $FH_{d}$. This is an unusual two-Higgs
doublet model where every Standard Model fermion $F$ couples to both
$H_{u}$ and $H_{d}$ (not their conjugates), which implies that in
some cases the remaining fermion in the interaction must be exotic.
This is not a problem since the extra fields can be made heavy via
the $m_{F}$ mass term in equation (\ref{eq:yukawa-interactions-extra}).

Let us consider in the following the lepton sector only. The decomposition
of the various representations is as follows:
\begin{align}
L & \rightarrow\underbrace{\left(\boldsymbol{2},-1/2\right)}_{\ell_{D}}\,,\\
\mathbb{L} & \rightarrow\left(\boldsymbol{3},0\right)+\left(\boldsymbol{3},-1\right)+\left(\boldsymbol{1},0\right)+\underbrace{\left(\boldsymbol{1},-1\right)}_{\ell_{S}}\,,\\
\overline{\mathbb{L}} & \rightarrow\left(\boldsymbol{3},0\right)+\left(\boldsymbol{3},1\right)+\left(\boldsymbol{1},0\right)+\underbrace{\left(\boldsymbol{1},1\right)}_{\ell_{S}^{c\prime}}\,,\\
e^{c} & \rightarrow\underbrace{\left(\boldsymbol{1},1\right)}_{\ell_{S}^{c}}\,,\\
\mathbb{e}^{c} & \rightarrow\left(\boldsymbol{2},3/2\right)+\underbrace{\left(\boldsymbol{2},1/2\right)}_{\ell_{D}^{c}}\,,\\
\overline{\mathbb{e}^{c}} & \rightarrow\left(\boldsymbol{2},-3/2\right)+\underbrace{\left(\boldsymbol{2},-1/2\right)}_{\ell_{D}^{\prime}}\,.
\end{align}
The Standard Model charged leptons are a mixture of the charged components
of the $SU(2)\times U(1)$ representations labeled above. The nomenclature
keeps track of lepton number, which is conserved (a superscript $c$
denotes an anti-lepton), and the subscripts indicate whether a field
is part of an $SU(2)$ singlet ($S$) or a doublet ($D$). Inserting
the vacuum expectation values of the scalars, and collecting the fermions
in the vectors $\Psi=\left(\ell_{D},\ell_{D}^{\prime},\ell_{S}\right)^{T}$
and $\Psi^{c}=\left(\ell_{S}^{c},\ell_{S}^{c\prime},\ell_{D}^{c}\right)^{T}$
we get the mass term
\begin{equation}
\Psi^{T}\left(\begin{array}{ccc}
0 & y_{L}^{\Omega}\left\langle H_{d}\right\rangle  & y_{LE}^{\chi}\left\langle \chi\right\rangle \\
y_{E}^{\Omega}\left\langle H_{d}\right\rangle  & 0 & m_{E}\\
y_{LE}^{\chi\prime}\left\langle \chi\right\rangle  & m_{L} & 0
\end{array}\right)\Psi^{c}\,.
\end{equation}
The VEV of the $H_{u}$ doublet contained in $\Omega$ does not appear
in this expression. Presuming that $m_{E}$ and $m_{L}$ are substantially
larger than the scalar vacuum expectation values, the last expression
implies that there are two heavy Dirac fermions, with masses $\approx m_{E}$
and $\approx m_{L}$, and a light one with a mass
\begin{equation}
m_{\textrm{light}}\approx\left(\frac{y_{E}^{\Omega}y_{LE}^{\chi}}{m_{L}}+\frac{y_{L}^{\Omega}y_{LE}^{\chi\prime}}{m_{E}}\right)\left\langle H_{d}\right\rangle \left\langle \chi\right\rangle \,.
\end{equation}
The left-handed part of this field is mostly composed of $\ell_{D}$,
with a small admixture of $\ell_{D}^{\prime}$ and $\ell_{S}$; the
right-handed part is mostly formed from $\ell_{S}^{c}$, but also
from $\ell_{S}^{c\prime}$ and $\ell_{D}^{c}$ (to a lesser extent):
\begin{align}
\ell & \approx\ell_{D}-\frac{y_{LE}^{\chi}\left\langle \chi\right\rangle }{m_{L}}\ell_{D}^{\prime}-\frac{y_{L}^{\Omega}\left\langle H_{d}\right\rangle }{m_{E}}\ell_{S}\,,\\
\ell^{c} & \approx\ell_{S}^{c}-\frac{y_{LE}^{\chi\prime}\left\langle \chi\right\rangle }{m_{E}}\ell_{S}^{c\prime}-\frac{y_{E}^{\Omega}\left\langle H_{d}\right\rangle }{m_{L}}\ell_{D}^{c}\,.
\end{align}

The situation is analogous for quarks: due to small admixtures with
the fields in the spinor representations of $SO(5)$, the Standard
Model fermions can interact through $\WOne_{1}^{\mu}$ with heavy
new fermions, the latter having exotic quantum numbers.

Note that lepton number is conserved again, and there is a total of
5 neutrinos and 4 anti-neutrinos (per generation), so we conclude
without further calculations that one neutrino is massless. As in
the previous model, an extra $\left(\boldsymbol{1},\boldsymbol{1},\boldsymbol{3},1\right)$
scalar solves the problem.

\section{\label{sec5:Arbitrary-gauge-bosons}Gauge bosons with arbitrary quantum
numbers}

Most research on extensions of the Standard Model group $G_{SM}=SU(3)_{C}\times SU(2)_{L}\times U(1)_{Y}$
is based on just a handful of groups, and there is little freedom
(if any) to change the way fermions transform under them. Fermion
masses are the main culprit: by having these particles transform under
some arbitrary representation of the gauge symmetry, it is likely
that fermions with exotic quantum numbers and which are chiral under
$G_{SM}$ will also be part of the model (see for instance \cite{Fonseca:2015aoa}).
This is a concern because such fields can have at most an electroweak
scale mass.

On the other hand, the simple exercise of picking pairs of right-
and left-handed Standard Model fermions yields a list of quantum numbers
for vectors fields which would have interesting phenomenology consequences
\cite{delAguila:2010mx,Biggio:2016wyy,Fonseca:2016jbm}. However,
just a few of them have been incorporated into fully-fledged models,
mostly because of the above difficulty. For example, while looking
at vectors fields that can mediate the neutrinoless beta decay of
a proton, the paper \cite{Fonseca:2016jbm} argued that ultraviolet-complete
models might be possible only for a few of them.

That assessment was too pessimistic. In the following, I will argue
that one can build viable models containing gauge bosons in arbitrary
representations of the Standard Model symmetry group. I will proceed
in two steps:
\begin{enumerate}
\item It is possible to show that for any representation $\boldsymbol{X}$
of the Standard Model group $G_{SM}$, there is always a group $G$
containing $G_{SM}$ in such a way that its adjoint representation
includes $\boldsymbol{X}$.
\item Fermions can be assigned to representations of $G$ such that only
the Standard Model ones remain massless right before eletroweak symmetry
is broken. This guarantees that exotic new fermions can be made heavy.
Furthermore, no gauge anomalies are generated. The vector boson mentioned
above can couple to Standard Model fermions.
\end{enumerate}

The first step involves group theory only. For the sake of argument,
consider the following scenario which works for any $\boldsymbol{X}$,\footnote{The argument actually fails when $\boldsymbol{X}$ is inert under
the full $SU(3)_{C}\times SU(2)_{L}\times U(1)_{Y}$. However, it
is well known that the trivial representation $\boldsymbol{X}=(\boldsymbol{1},\boldsymbol{1},0)$
is obtainable from extra $U(1)$ factors, for example.} even though it might not yield the smallest group $G$. Take $\boldsymbol{S}$
to be the trivial representation $(\boldsymbol{1},\boldsymbol{1},0)$
of $G_{SM}$, and $\boldsymbol{S^{\prime}}=(\boldsymbol{1},\boldsymbol{1},y)$
with $y$ equal to minus the total hypercharge of all components of
$\boldsymbol{X}$. In other words, $U(1)_{Y}$ acts on the reducible
representation $\boldsymbol{X}\oplus\boldsymbol{S^{\prime}}$ via
a traceless matrix. If $n$ is the dimension of $\boldsymbol{X}$,
then one can embed $G_{SM}$ in $SU(n+2)$ in such a way that the
adjoint representation of the latter contains $\boldsymbol{X}$. To
see it, note that there is an embedding under which the fundamental
representation $\boldsymbol{F}$ of $SU(n+2)$ decompose as
\begin{equation}
\boldsymbol{F}\rightarrow\boldsymbol{X}\oplus\boldsymbol{S}\oplus\boldsymbol{S^{\prime}}\,.\label{eq:F-branching}
\end{equation}
After all, $\boldsymbol{X}\oplus\boldsymbol{S}\oplus\boldsymbol{S^{\prime}}$
is represented by a set of twelve ($n+2$)-dimensional matrices which
are traceless and hermitian, hence they form a subalgebra of $SU(n+2)$.
Moving on to the adjoint representation, it transforms in the same
way as $\boldsymbol{F}\times\boldsymbol{F}^{*}$ with a singlet subtracted
(informally, we may express this as $\textrm{\textbf{Ad}}\sim\boldsymbol{F}\times\boldsymbol{F}^{*}-\boldsymbol{1}$),
so it follows directly from the previous branching rule that the adjoint
representation $\textrm{\textbf{Ad}}$ of $SU(n+2)$ decomposes as
\begin{equation}
\textrm{\textbf{Ad}}\rightarrow\boldsymbol{X}\oplus\boldsymbol{X}^{*}\oplus\textrm{`more'},\label{eq:Ad-branching}
\end{equation}
with $\textrm{`more'}=\left(\boldsymbol{X}\times\boldsymbol{X}^{*}\right)\oplus\boldsymbol{S^{\prime}}\oplus\boldsymbol{S}^{\boldsymbol{\prime}*}\oplus\left(\boldsymbol{X}\times\boldsymbol{S}^{\boldsymbol{\prime}*}\right)\oplus\left(\boldsymbol{X}^{*}\times\boldsymbol{S^{\prime}}\right)\oplus\boldsymbol{S}$.
As an example, we can infer immediately that a gauge boson with the
unusual quantum numbers $\boldsymbol{X}=\left(\boldsymbol{1},\boldsymbol{5},8\right)\equiv\boldsymbol{5}_{8}$
can be obtained from $SU(7)$ through the embedding defined by the
branching rules
\begin{align}
\boldsymbol{F}\equiv\boldsymbol{7} & \rightarrow\underbrace{\boldsymbol{5}_{8}}_{\boldsymbol{X}}\oplus\underbrace{\boldsymbol{1}_{0}}_{\boldsymbol{S}}\oplus\underbrace{\boldsymbol{1}_{-40}}_{\boldsymbol{S^{\prime}}}\,,\\
\textrm{\textbf{Ad}}\equiv\boldsymbol{48} & \rightarrow\underbrace{\boldsymbol{5}_{8}\oplus\boldsymbol{5}_{-8}}_{\boldsymbol{X}+\boldsymbol{X}^{*}}\oplus\underbrace{\boldsymbol{1}_{-40}\oplus\boldsymbol{1}_{40}}_{\boldsymbol{S^{\prime}}+\boldsymbol{S}^{\boldsymbol{\prime}*}}\oplus\underbrace{\boldsymbol{1}_{0}\oplus\boldsymbol{3}_{0}\oplus\boldsymbol{5}_{0}\oplus\boldsymbol{7}_{0}\oplus\boldsymbol{9}_{0}}_{\boldsymbol{X}\times\boldsymbol{X}^{*}}\oplus\underbrace{\boldsymbol{5}_{48}\oplus\boldsymbol{5}_{-48}}_{\boldsymbol{X}\times\boldsymbol{S}^{\boldsymbol{\prime}*}+\boldsymbol{X}^{*}\times\boldsymbol{S^{\prime}}}\oplus\underbrace{\boldsymbol{1}_{0}}_{\boldsymbol{S}}\,.
\end{align}

The above reasoning works for any $\boldsymbol{X}$, but it is unlikely
to involve the smallest possible group. For several quantum numbers
of the vector field, the reader can see in table 1 of \cite{Fonseca:2016jbm}
what are the minimal groups. To illustrate the point, $\boldsymbol{X}=\left(\boldsymbol{8},\boldsymbol{3},0\right)$
can be obtained from $SU(26)$ by the above argument, however, it
can also be extracted from the much smaller $SU(6)\times U(1)$ group.\footnote{It corresponds to the first branching rule (out of three) given by
the command \texttt{DecomposeRep{[}\{SU6,U1\}, Adjoint{[}\{SU6,U1\}{]},
\{SU3,SU2,U1\}{]}} in \texttt{GroupMath} \cite{Fonseca:2020vke}.}

~

Having settled this mathematical part of the problem, it remains to
be seen whether or not one can build realistic models based on the
above group embeddings. In principle, the solution adopted in this
work for the $\WOne_{1}$ vector field --- and which has also been
used in models for the B-anomalies \cite{Georgi:2016xhm,Diaz:2017lit,DiLuzio:2017vat,DiLuzio:2018zxy}
--- can be adapted to gauge bosons with other representations. We
may start by extending $G$ to $G\times SU(3)\times SU(2)\times U(1)$
(some of these factors, such as $SU(3)$ on the earlier models for
$\WOne_{1}$, might be unnecessary). The Standard Model symmetry group
$G_{SM}$ is obtained from the diagonal subgroup of an $SU(3)^{\prime}\times SU(2)^{\prime}\times U(1)^{\prime}$
contained in $G$, and $G_{321}\equiv SU(3)\times SU(2)\times U(1)$
outside it. The model will contain fermions $F=Q,u^{c},d^{c},L,e^{c}$
which transform as usual under $G_{321}$ and have a trivial $G$
charge. One must also add some scalars to correctly break the extended
gauge group and to couple to fermions via Yukawa interactions.

This is one possibility. There are no gauge anomalies and, with an
appropriate scalar sector, it should be feasible to obtain the Standard
Model as a low energy effective theory. Importantly, fermions will
not couple directly to the gauge bosons of $G$.

The situation changes if we introduce for each (or at least some)
$F$ a corresponding pair of vector-like fermions $\left(\mathbb{F},\overline{\mathbb{F}}\right)$
transforming non-trivially under the group $G$, and with the same
$G_{321}$ quantum numbers as $F$. (In the case of $F=Q$ it might
be convenient to add two pairs of vector-like fermions, as explained
shortly.) The recipe can be as follows for a gauge boson in some representation
$\boldsymbol{X}$ of $G_{SM}$:
\begin{enumerate}
\item Find a group $G$ whose adjoint representation contains $\boldsymbol{X}$.
This requirement can always be fulfilled. The full symmetry of the
model shall be given by the $G\times G_{321}$ group.
\item Pick a non-trivial representation $\boldsymbol{R}$ of $G$ such that
$\left(\boldsymbol{R},\boldsymbol{1},\boldsymbol{1},0\right)$ of
$G\times G_{321}$ contains the Standard Model sub-representation
$\left(\boldsymbol{1},\boldsymbol{2},-1/2\right)$.
\item Introduce two scalars $\Omega=\left(\boldsymbol{R},\boldsymbol{1},\boldsymbol{1},0\right)$
and $\chi=\left(\overline{\boldsymbol{R}},\boldsymbol{1},\boldsymbol{2},-1/2\right)$;
they include at least one Higgs doublet $\left(\boldsymbol{1},\boldsymbol{2},-1/2\right)$
and a singlet $\left(\boldsymbol{1},\boldsymbol{1},0\right)$.\footnote{If the VEV of this last field is insufficient to correctly break $G\times G_{321}$
down to the Standard Model group, one must add more scalars.}
\item Introduce Weyl fermions $F=Q,u^{c},d^{c},L,e^{c}$ transforming as
$\left(\boldsymbol{1},\boldsymbol{3},\boldsymbol{2},1/6\right)$,
$\left(\boldsymbol{1},\overline{\boldsymbol{3}},\boldsymbol{1},-2/3\right)$,
$\left(\boldsymbol{1},\overline{\boldsymbol{3}},\boldsymbol{1},1/3\right)$,
$\left(\boldsymbol{1},\boldsymbol{1},\boldsymbol{2},-1/2\right)$
and $\left(\boldsymbol{1},\boldsymbol{1},\boldsymbol{1},1\right)$.
For each $F$ we need a vector-like fermion pair $\left(\mathbb{F},\overline{\mathbb{F}}\right)$
such that $\mathbb{F}$ transforms as $F$ under $G_{321}$ and as
a $\boldsymbol{R}$ under $G$. However, since the $\boldsymbol{R}$
representation contains only a down-like Higgs doublet, $\left(\boldsymbol{R},\boldsymbol{1},\boldsymbol{1},0\right)\rightarrow\left(\boldsymbol{1},\boldsymbol{2},-1/2\right)+\cdots$,
$\left(\mathbb{Q},\overline{\mathbb{Q}}\right)$ will contain only
down-like quarks --- $\left(\overline{\boldsymbol{3}},\boldsymbol{1},1/3\right)$
and $\left(\boldsymbol{3},\boldsymbol{1},1/3\right)$. In order to
treat all quarks equally, we may want to introduce two vector-like
fermions in association to $F=Q$:
\begin{equation}
\mathbb{Q}_{u}\equiv\left(\overline{\boldsymbol{R}},\boldsymbol{3},\boldsymbol{2},1/6\right)\textrm{ and }\mathbb{Q}_{d}\equiv\left(\boldsymbol{R},\boldsymbol{3},\boldsymbol{2},1/6\right)\,.
\end{equation}
This is not needed if $\boldsymbol{R}$ contains both $\left(\boldsymbol{1},\boldsymbol{2},-1/2\right)$
and $\left(\boldsymbol{1},\boldsymbol{2},1/2\right)$ (as in the $SO(5)$
models of sections \ref{sec3:A-minimal-model} and \ref{sec4:An-alternative:-charging}).
\item If $\boldsymbol{R}$ is complex, the list of fermion masses and Yukawa
terms is the following:
\begin{align}
\textrm{masses: } & \mathbb{Q}_{u}\overline{\mathbb{Q}_{u}}\,,\;\mathbb{Q}_{d}\overline{\mathbb{Q}_{d}}\,,\;\mathbb{u}^{c}\overline{\mathbb{u}^{c}}\,,\;\mathbb{d}^{c}\overline{\mathbb{d}^{c}}\,,\;\mathbb{L}\overline{\mathbb{L}}\,,\;\mathbb{e}^{c}\overline{\mathbb{e}^{c}}\,,\\
\textrm{singlet interactions: } & Q\mathbb{u}^{c}\chi^{*}\,,\;\mathbb{Q}_{u}u^{c}\chi^{*}\,,\;Q\mathbb{d}^{c}\chi\,,\;\mathbb{Q}_{d}d^{c}\chi\,,\;L\mathbb{e}^{c}\chi\,,\;\mathbb{L}e^{c}\chi\,,\\
\textrm{doublet interactions: } & Q\overline{\mathbb{Q}_{u}}\Omega^{*}\,,\;u^{c}\overline{\mathbb{u}^{c}}\Omega^{*}\,,\;Q\overline{\mathbb{Q}_{d}}\Omega\,,\;d^{c}\overline{\mathbb{d}^{c}}\Omega\,,\;L\overline{\mathbb{L}}\Omega\,,\;e^{c}\overline{\mathbb{e}^{c}}\Omega\,.
\end{align}
By design all but the Standard Model fermions can be made heavy without
any tuning. Assuming that the scalar VEVs are smaller than the vector-like
masses $m_{F}\mathbb{F}\overline{\mathbb{F}}$, the light fermion
mass eigenstates are composed mostly of the $F$'s (that is $Q,u^{c},d^{c},L$
and $e^{c}$). Mostly, but not entirely: a particularly important
consequence is that through mixing the Standard Model fermions will
couple to the gauge bosons of $G$. Furthermore, note that the VEV
of $\chi$ does not break $G_{SM}$, thus it can be comparable (or
even greater than) the masses $m_{F}$; as a consequence, fermion
mixing might be large.\footnote{There is a caveat. For every $\boldsymbol{X}$, it is always possible
to pick a $G$ and an $\boldsymbol{R}$ fulfilling steps 1 and 2.
However, by itself this does not ensure that the gauge bosons of $G$
transforming as $\boldsymbol{X}$ will couple to the important sub-representations
in the $\mathbb{F}/\overline{\mathbb{F}}$ fields (i.e., those which
can mix with the $F$'s). This should not be a concern as long as
$\boldsymbol{X}$, $G$ and $\boldsymbol{R}$ and not too exotic.
However, in general, one has an extra requirement which --- it turns
out again --- can always be met. I argued that an $n$-dimensional
$\boldsymbol{X}$ is contained in the adjoint representation of $SU(n+2)$,
as shown by the branching rules (\ref{eq:F-branching}) and (\ref{eq:Ad-branching}).
Those decompositions do not ensure, as we would like, that the down
Higgs doublet $H_{d}$ in $\Omega$ couples to anything else through
$\boldsymbol{X}$. We can fix that in $SU(n+3)$ with $\boldsymbol{R}=\textrm{fundamental rep.}\rightarrow\boldsymbol{H_{d}}\oplus\boldsymbol{X}^{\prime}\oplus\boldsymbol{S^{\prime}}$
where $\boldsymbol{X}^{\prime}$ is a representation in the product
$\boldsymbol{H_{d}}\times\boldsymbol{X}^{*}$ and, as before, $\boldsymbol{S^{\prime}}$
makes $\boldsymbol{H_{d}}\oplus\boldsymbol{X^{\prime}}\oplus\boldsymbol{S^{\prime}}$
traceless under $U(1)_{Y}$. With this choice of group embedding,
the $H_{d}$ scalar in $\Omega$ couples to something else ($\boldsymbol{X}^{\boldsymbol{\prime}}$)
via the gauge bosons transforming as $\boldsymbol{X}$, and so do
the other important components in $\chi$ and the $\mathbb{F}/\overline{\mathbb{F}}$
fermions.}
\end{enumerate}
An analogous list of interactions can be compiled when $\boldsymbol{R}$
is (pseudo)real. In that case there is no need for both $\mathbb{Q}_{u}$
and $\mathbb{Q}_{d}$ (a single vector-like $\mathbb{Q}$ is sufficient).

~

Lastly, it is interesting that baryon and lepton numbers are preserved
in this construction. That was also the case in the $SO(5)$ models
for $\WOne_{1}$, where neutrinos are Dirac particles even though
there are right-handed neutrinos and scalars which are capable of
inducing Majorana masses.

As a further example, consider a gauge boson $X_{\mu}$ with the quantum
numbers $\boldsymbol{X}=\left(\boldsymbol{3},\boldsymbol{2},5/6\right)$.
The argument reported earlier points to the $SU(8)$ group, but the
field can also be obtained in the widely studied $SU(5)$ model of
grand unification \cite{Georgi:1974sy}. $X_{\mu}$ induces proton
decay via its simultaneous coupling to the Standard Model bilinears
$Q\overline{u^{c}}$ and $L\overline{e^{c}}$, hence it must be an
extremely heavy field. An alternative is to forbid one of its two
problematic couplings with a symmetry that, for example, enforces
baryon-number conservation. Given the stringent limits on the proton's
lifetime \cite{Super-Kamiokande:2016exg}, a $X_{\mu}$ field at the
TeV scale is problematic even if it induces nucleon decay through
loops only.

Now consider an $SU(5)\times SU(3)\times SU(2)\times U(1)$ model,
with $\boldsymbol{R}=\boldsymbol{5}$. Despite the complexity of the
list of fermion masses and interactions (see above), one can a assign
an unbroken baryon number $B$ to all fields: $B\left(Q,\mathbb{Q}_{u/d},\overline{\mathbb{u}^{c}},\overline{\mathbb{d}^{c}}\right)=-B\left(\overline{\mathbb{Q}_{u/d}},u^{c},\mathbb{u}^{c},d^{c},\mathbb{d}^{c}\right)=1/3$.
The same holds for lepton number. A model constructed along these
lines should therefore predict a stable proton, even if the $SU(5)\times SU(3)\times SU(2)\times U(1)$
symmetry breaking scale is as low as a few TeV. This is true also
for other groups and other $\boldsymbol{X}$'s: more fields are needed
in order to break baryon and/or lepton number.

\section{\label{sec:Final-comments}Summary}

A vector field $\WOne_{1}$ with the quantum numbers $\left(\boldsymbol{3},1\right)$
under $SU\left(2\right)_{L}\times U(1)_{Y}$ cannot couple to pairs
of Standard Model fermions, yet in principle it could interact with
two Higgs doublets. While this is true, I have argued in this paper
that such coupling will not be generated in a Yang-Mills theory, if
$\WOne_{1}$ is a gauge boson. As a consequence, the suggestion in
\cite{Bagnaschi:2022whn,DiLuzio:2022xns} that such a field could
explain the recent CDF measurement of the $W$-boson mass becomes
less appealing.

Notwithstanding the lack of the above interaction, in this paper I
considered a minimal model containing $\WOne_{1}$ as a gauge boson,
potentially with a TeV scale mass. In fact, I considered two closely
related models: one where this field does not interact with fermions
at all, and another where it does. In the latter, due to its quantum
numbers, the $\WOne_{1}$ coupling to Standard Model fermions involves
necessarily exotic ones as well. The two models are based on an $SO(5)\times SU(2)\times U(1)$
extended electroweak group, and therefore they predict the existence
of a $W^{\prime}$ and a $Z^{\prime}$, with a mass hierarchy $m_{\WOne_{1}}<m_{Z^{\prime}}<m_{W^{\prime}}$.
Incidentally, these two vector fields are known to affect the $W$
mass (see \cite{Strumia:2022qkt,DiLuzio:2022xns,Bagnaschi:2022whn,Thomas:2022gib,Cheng:2022aau,Cai:2022cti,Zhang:2022nnh}):
$W^{\prime}$ pulls it down and $Z^{\prime}$ has the opposite effect.
Since the $Z^{\prime}$ is lighter, it could in principle explain
the CDF data. The two models also contain new scalars (shown in tables
\ref{tab:scalars} and \ref{tab:model-2}).

Finally, in the last part of this work I have argued that it is possible
to extend the argument used here for $\WOne_{1}$ --- and elsewhere
for the $U_{\mu}$ lepto-quark \cite{Georgi:2016xhm,Diaz:2017lit,DiLuzio:2017vat,DiLuzio:2018zxy}
--- to gauge bosons with arbitrary quantum numbers. Some phenomenological
limitations do apply: for example, models with a colorless and fractionally
charged field contain necessarily a stable electrically charged particle,
which is a problem in astrophysics and cosmology. However, even with
this kind of consideration, many viable possibilities remain, and
therefore the existence of TeV-scale gauge bosons with a wide variety
of quantum numbers cannot be ruled out.

\section*{Acknowledgments}

I discussed aspects of this paper with several colleagues, to whom
I'm indebted. In particular, I would like to thank Mikael Chala, José
Ignácio Illana and Manuel Pérez-Victoria for fielding many of my questions,
as well as Supratim Bakshi, Jorge de Blas, Javi Fuentes, and José
Santiago. Furthermore, I acknowledge the financial support from MCIN/AEI
(10.13039/501100011033) through grant number PID2019-106087GB-C22
and from the Junta de Andalucía through grant number P18-FR-4314 (FEDER).

\end{document}